\documentstyle[epsf]{article}
\begin{document}

 \

 \

\begin{center}
{\Large \bf 
Compositeness of gauge boson and asymptotic freedom in non-abelian 
gauge theory
}
\end{center}

 \

 \

 \

 \

 \

\begin{center}
Takashi Hattori \footnote{ e-mail : hattorit@kdcnet.ac.jp }
\end{center}

\begin{center}
{\it
Department of Physics, Kanagawa Dental College, 
Inaoka-cho, Yokosuka, Kanagawa, 238-8580, Japan
}
\end{center}

 \

 \

 \

 \

 \

 \

 \

\begin{abstract}
In order to investigate the composite gauge field,
 we consider the compositeness condition
 (i.e. renormalization constant $Z_3=0$) in the
 general non-abelian gauge field theory.
We calculate $Z_3$ at the next-to-leading order in
 $1/N_f$ expansion
 ($N_f$ is the number of fermion flavors), and obtain the
 expression to the gauge coupling constant through the
 compositeness condition.
Then the gauge coupling constant is proportional to
 $1/\sqrt{4N_f T(R)-11C_2(G)}$
 where $T(R)$ is the index for a representation $R$ of gauge
 group $G$, and $C_2(G)$ is the quadratic Casimir.
It is found that the gauge boson compositeness take place only when
 $N_f T(R)/C_2(G) > 11/4$, in which the asymptotic freedom
 in the non-abelian gauge field theory fails.
\end{abstract}

\newpage

\section{ INTRODUCTION }

The compositeness of vector boson \cite{cg}--\cite{AH} is a interesting
 subject in present particle physics.
The idea of composite gauge boson has been applied in
 quark-lepton physics \cite{qlphys} and hadron physics \cite{hadphys}.
The experimental results \cite{exp} do not exclude the possibility of
 compositeness of gauge boson.
In the composite gauge field theory, the bare Lagrangian does not
 include explicitly the kinetic term of gauge field.
The kinetic term of gauge field, however, be derived from the
 quantum fluctuations of matter field, in which the gauge field
 becomes consequently dynamical one, and then the gauge boson is
 regarded as a composite which is composed of the matter fields.
When we consider the higher-order contributions of the model,
 tremendous divergences prevent us to solve perturbatively
 the problem of composite state.
In order to obtain physical predictions, people use the ladder
 approximation or the renormalization group equation.
In these method, the contributions from particular diagrams are calculated
 through the expansion with gauge coupling constant.
For a solution of the problem of composite state,
 we use the gauge theory with the compositeness condition \cite{ccg},
 and then consider the contributions from next-to-leading diagrams
 through the $1/N_f$ expansion where $N_f$ is the number of
 matter-fermion flavors.
In general, the gauge field theory with the compositeness condition $Z_3=0$
  is reduced to the composite gauge field theory with the finite
 cutoff $\Lambda$ (i.e. new physics scale) where $Z_3$ is the wave
 function renormalization constant of the gauge field.
The gauge field theory with $Z_3=0$ provides a useful method to
 investigate the composite gauge field when the physical cutoff scale
 $\Lambda$ is very large than the present energy scale.
Hence we calculate the renormalization constant $Z_3$
 with the $1/N_f$ expansion within the gauge field theory.
Then the information of composite gauge field theory can be
 obtained from the gauge field theory with $Z_3=0$.

In the previous papers \cite{AH},
 K. Akama and the present author investigated the compositeness
 condition in abelian ($U(1)$) and non-abelian ($SU(N_c)$) gauge
 field theories at next-to-leading order $1/N_f$.
The compositeness condition in Nambu-Jona-Lasinio model \cite{NJL}
 at the next-to-leading order was investigated in Ref. \cite{cA}.

In this paper, we present a detailed formulation of the scheme for
 the compositeness condition of general non-abelian gauge field, and
 discuss {\it a complementarity} between gauge boson compositeness
 and asymptotic freedom \cite{af} in the gauge field theory.
In Sec. II, we present the composite gauge field
 which is composed of matter fermions,
 and then discuss the gauge field theory with
 compositeness condition ($Z_3=0$).
In Sec. III, we calculate the renormalization constant $Z_3$ to the
 next-to-leading order in $1/N_f$ expansion \cite{AH} within the
 general non-abelian gauge field theory.
In Sec. IV, we obtain the expression to the gauge coupling constant
 through the compositeness condition.
In Sec. V, we give the conclusions of this paper.

\section{ COMPOSITENESS AND RENORMALIZATION }
\subsection{Compositeness condition}

We consider the composite gauge field theory with $N_f$ matter
 fermions
 $\psi^j$ ($j=1,2, \cdots , N_f$) each of which
 belong to the representation of color gauge group $G$.
The Lagrangian is
 given by the following gauge invariant form without the
 kinetic term of bare gauge field,
\begin{eqnarray}
 {\cal L}_A
 =
 \overline \psi^j \left( i \! \not \! \partial - m
 + T^a \! \! \not \! \! A^a \right) \psi^j ,
\label{LA}
\end{eqnarray}
 where $T^a$ is the generator of gauge group $G$, and
 $m$ is the bare mass of matter fermions $\psi^j$.
For example, if we consider the matter fermions $\psi^j$ which
 belong to the fundamental representation of $SU(N_c)$ group,
 then $T^a$ is given by $T^a=\lambda^a/2$
 ($a=1,2, \cdots, N_c^2-1$),
 where $\lambda^a$ are the $SU(N_c)$ Gell-Mann matrices.
In (\ref{LA}), the fields $A^a_\mu$ may be regarded as auxiliary
 field without independent dynamical degree of freedom,
 because if we use the Euler equation with respect to
 $A^a_{\mu}$ for (\ref{LA}), then
 $\partial {\cal L}_A/\partial A^a_{\mu} = 0$ give
 the constraint condition with respect to $A^a_{\mu}$,
 as follows
\begin{eqnarray}
  \overline \psi^j T^a \gamma_\mu \psi^j =0.
\label{EULA}
\end{eqnarray}
The kinetic term of non-abelian gauge field is induced through
 quantum fluctuation of the fermion fields $\psi^j$,
 and hence the gauge field becomes dynamical one.
This implies that the gauge boson is a composite made of a pair of
 the fermion $\psi^j$ and its anti-particle.
We note that (\ref{LA}) corresponds to strong coupling limit
 $(F \rightarrow \infty)$ of the following Nambu-Jona-Lasinio
 type model,
 where $F$ is the coupling constant
\begin{eqnarray}
 {\cal L}_N
 = \overline \psi^j
 \left( i \! \not \! \partial - m \right) \psi^j
 + F \left( \overline \psi^j T^a \gamma_\mu \psi^j \right)^2
\label{LNF}
\end{eqnarray}
 which is equivalent to
\begin{eqnarray}
 {\cal L}_A'
 = \overline \psi^j \left( i \! \not \! \partial - m \right) \psi^j
 + \overline \psi^j T^a \! \! \not \! \! A^a \psi^j
 -{1 \over 4F}(A^a_\mu)^2.
\label{LN}
\end{eqnarray}
The theory (\ref{LA}) involves severe ultraviolet divergences.
In order to absorb the divergences in part, we rescale
 the fields and the mass as
\begin{eqnarray}
 \psi^j = \sqrt{X_\psi} \psi^j_{\rm r}, \ \ \ \ \
 A^a_\mu= \sqrt{X_A} G^a_{{\rm r} \mu}, \ \ \ \ \
 m = {X_m \over X_\psi} m_{\rm r} ,
\label{DA}
\end{eqnarray}
 then (\ref{LA}) is rewritten as
\begin{eqnarray}
 {\cal L}_A
 =
 X_\psi \overline \psi^j_{\rm r}
 \left( i \! \not \! \partial - {X_m \over X_\psi} m_{\rm r} \right)
 \psi^j_{\rm r}
 + X_\psi \sqrt{X_A} \ \ \overline \psi^j_{\rm r}
 T^a \not \! \! G^a_{\rm r} \psi^j_{\rm r} ,
\label{LAR}
\end{eqnarray}
 where $X_\psi$, $X_A$, and $X_m$ are rescaling factors
 for the bare quantities $\psi^j$, $A^a_\mu$, and $m$,
 respectively.

Now we see that the composite gauge field theory (\ref{LA}) is the special case of
 the ordinary gauge field theory specified by the compositeness
 condition $Z_3=0$.
The non-abelian gauge theory with gauge field $G^a_\mu$ and
 fermionic fields $\psi^j$ is given by the Lagrangian
\begin{eqnarray}
 {\cal L}_G
 =
 -{1 \over 4} \left( G^a_{ \mu \nu} \right )^2
 + \overline \psi^j
 \left( i \! \not \! \partial - m + g T^a \! \! \not \! \! G^a
 \right) \psi^j
\label{LG}
\end{eqnarray}
 with
 $G^a_{\mu \nu} =
 \partial_\mu G^a_\nu - \partial_\nu G^a_\mu
 + g f^{abc} G^b_\mu G^c_\nu $,
 where $g$ is the gauge coupling constant, and $f^{abc}$ is
 the structure constant of gauge group $G$.
In order to absorb the ultraviolet divergences in quantum
 corrections,
 we renormalize the bare quantities in (\ref{LG}) as
\begin{eqnarray}
 \psi^j = \sqrt{Z_2} \psi^j_{\rm r}, \ \ \ \ \
 G^a_\mu = \sqrt{Z_3} G^a_{{\rm r} \mu}, \ \ \ \ \
 g = {Z_1 \over Z_2 \sqrt{Z_3}} g_{\rm r}, \ \ \ \ \
 m = {Z_m \over Z_2} m_{\rm r} ,
\label{DG}
\end{eqnarray}
 then (\ref{LG}) is rewritten as
\begin{eqnarray}
 {\cal L}_G
=
 -{1 \over 4} Z_3 \left( G^a_{{\rm r} \mu \nu} \right)^2
 + Z_2 \overline \psi^j_{\rm r}
 \left( i \! \not \! \partial - {Z_m \over Z_2} m_{\rm r}
 + {Z_1 \over Z_2} g_{\rm r} T^a \! \! \not \! \! G^a_{\rm r}
 \right) \psi^j_{\rm r}
\label{LGR}
\end{eqnarray}
 with
 $G^a_{{\rm r} \mu \nu}=
 \partial_\mu G^a_{{\rm r} \nu}
 - \partial_\nu G^a_{{\rm r} \mu}
 +(Z_1 / Z_2) g_{\rm r} f^{abc} G^b_{{\rm r} \mu}
 G^c_{{\rm r} \nu}$,
 where the quantities with the subscript "r" are
 the renormalized ones,
 and $Z_1$, $Z_2$, $Z_3$, and $Z_m$ are the renormalization
 constants.
When we use the following copmositeness condition for the
 non-abelian gauge field,
\begin{eqnarray}
 Z_3=0,
\label{COMP}
\end{eqnarray}
 (\ref{LGR}) reduces to
\begin{eqnarray}
 {\cal L}_G
 =
 Z_2 \overline \psi^j_{\rm r}
 \left( i \! \not \! \partial - {Z_m \over Z_2} m_{\rm r}
 + {Z_1 \over Z_2} g_{\rm r} T^a \! \! \not \! \! G^a_{\rm r}
 \right) \psi^j_{\rm r},
\label{LGRC}
\end{eqnarray}
 in which the kinetic term of non-abelian gauge field
 $G^a_{{\rm r} \mu}$ vanishes.
If we set the following equalities,
\begin{eqnarray}
 Z_2 = X_\psi, \ \ \ \ \
 {Z_m \over Z_2} = {X_m \over X_\psi}, \ \ \ \ \
  Z_1 g_{\rm r} = X_\psi \sqrt{X_A} ,
\label{EAG}
\end{eqnarray}
 then (\ref{LGRC}) is identical to (\ref{LAR}).
Hence the non-abelian gauge field theory (\ref{LG})
 with $Z_3=0$ is
 equivalent to the composite gauge field theory (\ref{LA}).

Hereafter we consider the gauge field theory (\ref{LG}) with
 compositeness condition ($Z_3=0$) in order to investigate
 the composite gauge field.
In the calculation of quantum effects, we need to fix the gauge
 to guarantee that the inversed gauge boson propagator exists.
Then the gauge fixing term and the Faddeev-Popov ghost term are
 introduced into (\ref{LG}) and (\ref{LGR}), as follows
\begin{eqnarray}
 {\cal L}_G^{{\rm g.f.}}
 &=&
 {\cal L}_G
 - {1 \over 2 \xi} \left( \partial^\mu G^a_\mu \right)^2
 + \partial^\mu \eta^{a \dagger}
 \left( \partial_\mu \eta^a - g f^{abc} \eta^b G^c_\mu \right)
\cr
 &=&
 {\cal L}_G
 - {1 \over 2 \xi_{\rm r}}
 \left( \partial^\mu G^a_{{\rm r} \mu } \right)^2
 + Z_\eta \partial^\mu \eta^{a \dagger}_{\rm r}
 \left( \partial_\mu \eta^a_{\rm r}
 - {Z_1 \over Z_2} g_{\rm r} f^{abc} \eta^b_{\rm r}G^c_{{\rm r} \mu}
 \right) ,
\label{LGFIX}
\end{eqnarray}
 where $\xi$ is the gauge fixing parameter, and
 $\eta^a \ (a=1,2, \cdots, N_c^2-1)$ is the Faddeev-Popov
 ghost field,
 and $Z_\eta$ is the renormalization constant for $\eta^a$.
The gauge fixing in (\ref{LGFIX}) dos not alter the physical
 quantities,
 and hence (\ref{LGFIX}) is equivalent to (\ref{LG})
 and (\ref{LGR}) \cite{Muta}.

\subsection{Leading order contribution to $Z_3$}

In order to find $Z_3$, we use rather $1/N_f$ expansion
 than usual
 perturbation expansion with the coupling constant $g_{\rm r}$,
 because the latter expansion fails under the
 compositeness condition ($Z_3=0$) \cite{AH}.
At the leading order in $1/N_f$ expansion, $Z_3$ is chosen so as to
 cancel the divergence to the one-fermion-loop
 diagram A in Fig.\ref{FigOne},
 where the solid and wavy lines indicate the fermion and
 the gauge boson propagators, respectively.
In the dimensional regularization, the one-loop vacuum polarization
 tensor is given by
\begin{eqnarray}
 \Pi^{ab \ {\rm A}}_{\mu \nu}(p)
 =
 {4 \over 3}{1 \over (4\pi)^2} N_f g_{\rm r}^2  \Gamma (\epsilon)
 {1 \over (-p^2)^\epsilon }
 \left( -g_{\mu \nu} p^2 + p_\mu p_\nu \right)
  {\rm Tr}(T^a T^b)
  + O(m_{\rm r}^2) ,
\label{A}
\end{eqnarray}
 where $p_\mu$ is the momentum of external gauge boson line,
 and $\epsilon$ is given by
 $\epsilon=(4-d)/2$ with the number of space-time
 dimensions $d$.
Then the renormalized tensor
 $\Pi^{ab \ {\rm A}}_{\mu \nu\ {\rm ren}}(p)$ is given by
\begin{eqnarray}
 \Pi^{ab \ {\rm A}}_{\mu \nu\ {\rm ren}}(p)
 = \Pi^{ab \ {\rm A}}_{\mu \nu}(p)
 +(Z_3 - 1)
 \left(-g_{\mu \nu} p^2 + p_\mu p_\nu \right) \delta^{ab}.
\label{Ar}
\end{eqnarray}
The leading divergent part
 $\Pi^{ab \ {\rm A}}_{\mu \nu\ {\rm div}}(p)$ is expressed as
\begin{eqnarray}
 \Pi^{ab \ {\rm A}}_{\mu \nu\ {\rm div}}(p)
 =
 {4 \over 3} N_f T(R) g_{\rm r}^2 I
 \left( -g_{\mu \nu} p^2 + p_\mu p_\nu \right) \delta^{ab}
\label{Adiv}
\end{eqnarray}
 with $T(R) \delta^{ab}={\rm Tr}(T^a T^b)$,
 where $I$ is given by
\begin{eqnarray}
 I
 =
 {1 \over 16 \pi^2 } \Gamma(\epsilon) \approx
 {1 \over 16 \pi^2 \epsilon }.
\label{I}
\end{eqnarray}
If we use the Pauli-Villars regularization, $I$ is given by
\begin{eqnarray}
 I
 =
 {1 \over 16 \pi^2 } {\rm ln} \left({\Lambda^2 \over m_{\rm r}^2} \right) ,
\label{IPV}
\end{eqnarray}
 where $\Lambda$ is the momentum cutoff.
In order to cancel the divergence, we choose $Z_3$ in (\ref{Ar}),
 as follows
\begin{eqnarray}
 Z_3 = 1 - {4 \over 3} N_f T(R) g_{\rm r}^2 I.
\label{Z0L}
\end{eqnarray}
For the renormalization constant $Z_3$,
 (\ref{Z0L}) gives the contribution at the leading order
 $O(1)$ in $1/N_f$ expansion.

When we use the compositeness condition (\ref{COMP})
 to (\ref{Z0L}),
 the gauge coupling constant $g_{\rm r}$ is expressed as
\begin{eqnarray}
 g_{\rm r}^2
 =
 {3 \over 4 N_f T(R)I } \propto {\epsilon \over N_f}.
\label{g0L}
\end{eqnarray}
For the gauge coupling constant $g_{\rm r}$,
 (\ref{g0L}) gives the result at the
 leading order $O(N_f^{-1})$ in $1/N_f$ expansion.
Notice that (\ref{g0L}) coincides with the known result
 of the composite gauge field theory \cite{cg}--\cite{qlphys}.

We take the regularization scheme as an approximation to some
 physical momentum cutoff $\Lambda$
 which implies the new physics scale.
Then the parameter $\epsilon$ should be fixed at the
 non-vanishing value
 from (\ref{I}) and (\ref{IPV}), as follows
\begin{eqnarray}
 \epsilon = {1 \over {\rm ln}(\Lambda^2 / m_{\rm r}^2)}.
\label{CUTOFF}
\end{eqnarray}
In (\ref{g0L}), we notice that the gauge coupling constant
 $g_{\rm r}$ is proportional to $ \sqrt{\epsilon / N_f}$.

\section{ NEXT-TO-LEADING CONTRIBUTIONS TO $Z_3$ }
\subsection{ Leading divergences}

We consider the next-to-leading contributions to
 renormalization constant $Z_3$ in $1/N_f$ expansion.
The diagrams are classified by powers of $1/ N_f$ instead of
 powers of the coupling constant $g_{\rm r}$.
In Fig.\ref{FigOne} (B--H), we show the diagrams for the next-to-leading
 contribution in the $1/N_f$ expansion, where the disks
 stand for
 insertions of an arbitrary number of one-fermion loops into
 the gauge boson propagator.
In addition to the one-boson-loop diagrams B and C, the
 multi-loop diagrams (D--H) belong to this order,
 because the gauge boson propagator with the fermion-loops
 inserted has the same order in $N_f^{-1}$ as
 the usual gauge boson propagator.
Though the integral of an $n$-loop diagram diverges like
 $(1/\epsilon)^n$, the diagram is suppressed by the factor
 $g_{\rm r}^{2n} \propto \epsilon^n$ from (\ref{g0L}).
Therefore, all the diagrams in Fig.\ref{FigOne} behave like $O(\epsilon^0)$
 at least.
Hereafter we return only the $O(\epsilon^0)$ contributions,
 which implies that we retain only the leading divergences.

The contribution from diagram A in Fig.\ref{FigOne} was discussed already
 in the previous
 section in which the diagram gives the result of leading
 order in $1/N_f$ expansion.
The contribution from diagram B is given by
\begin{eqnarray}
 \Pi^{ab \ {\rm B}}_{\mu \nu}
 =
 {g_{\rm r}^2 \over 16\pi^2} \Gamma (\epsilon)
 {1 \over (-p^2)^\epsilon}
 \left[{25 \over 12} g_{\mu \nu} p^2-{7 \over 3}p_\mu p_\nu
 -{1 \over 2} \xi_{\rm r}
 \left(g_{\mu \nu} p^2-p_\mu p_\nu \right) \right]
 C_2(G) \delta^{ab}
\label{BR}
\end{eqnarray}
 with $C_2(G) \delta^{ab}=f^{acd}f^{bcd}$.
The contribution from the Faddeev-Popov-ghost loop diagram C in Fig.\ref{FigOne} is
 given by
\begin{eqnarray}
 \Pi^{ab \ {\rm C}}_{\mu \nu}
 =
 -{g_{\rm r}^2 \over 16\pi^2} \Gamma (\epsilon -1)
 {1 \over (-p^2)^\epsilon}
 \left({1 \over 12} g_{\mu \nu} p^2
 + {1 \over 6}p_\mu p_\nu \right) C_2(G) \delta^{ab}.
\label{CR}
\end{eqnarray}

Now we discuss the each contribution from multi-loop
 diagrams D-H in Fig.\ref{FigOne}.
First let us present the gauge boson propagator
 $D^{ab \ [l]}_{\mu \nu}(p)$ into which
 an arbitrary number of one-fermion loops are inserted,
 where superscript $l$ is the number of one-fermion loops
 which are inserted into the gauge boson propagator.
The gauge boson propagator
 $D^{ab \ [l]}_{\mu \nu}(p)$ is defined as
\begin{eqnarray}
 D^{ab \ [l]}_{\mu \nu}(p)
 =
 D_{\mu \mu_1}^{aa_1}(p)\ \Pi_{a_1b_1}^{\mu_1 \nu_1}(p)\
 D_{ \nu_1 \mu_2}^{b_1a_2}(p)\ \Pi_{a_2b_2}^{\mu_2 \nu_2}(p)\
 \cdots \
 \Pi_{a_l b_l}^{\mu_l \nu_l}(p)\ D_{ \nu_l \nu}^{b_l b}(p) ,
\label{GPD}
\end{eqnarray}
 where
 $D_{\mu \mu_1}^{aa_1}(p)$, etc. are
 the usual gauge boson propagators,
 and $\Pi_{a_1b_1}^{\mu_1 \nu_1}(p)$, etc.
 are given by (\ref{A}).
Then we obtain the gauge boson propagator $D^{ab \ [l]}_{\mu \nu}(p)$,
 as follows
\begin{eqnarray}
 D^{ab \ [l]}_{\mu \nu}(p)
 =
  {1 \over (-p^2)^{1+\epsilon l} }
  \left(- {4 \over 3} N_f T(R) g_{\rm r}^2 I \right)^l
  \left[-g_{\mu \nu}+(1-\xi_{\rm r} \delta_{0 l})
  {p_\mu p_\nu \over p^2} \right] \delta^{ab}
\label{GPN}
\end{eqnarray}
 with $I= \Gamma (\epsilon)/16\pi^2 \approx 1/(16\pi^2 \epsilon) $,
 where $\delta_{0 l}$ is Kronecker's symbol.
For $l=0$, (\ref{GPN}) is reduced to the usual
 gauge boson propagator $D_{\mu \nu}^{ab}(p)$.

The contribution from diagram D is given by
\begin{eqnarray}
 \Pi^{ab \ {\rm D}}_{\mu \nu}(p)
 =
 -N_f \int {d^n k \over i(2\pi)^n } {\rm Tr}
 \left[ {1 \over m_{\rm r} -\not \! k}
 J(k)
 {1 \over m_{\rm r}-\not \! k}
 \gamma_\mu \left(g_{\rm r}T^a \right)
 {1 \over m_{\rm r}-(\not \! p + \not \! k)}
 \gamma_\nu \left(g_{\rm r}T^b \right) \right]
\label{D}
\end{eqnarray}
 with the fermion self-energy part
\begin{eqnarray}
 J(k)
 =
 -\int {dq^n \over i(2\pi)^n }
 \left[ \gamma^\rho \left(g_{\rm r}T^c \right)
 {1 \over m_{\rm r} -(\not \! k - \not \! q)} \gamma^\sigma
 \left(g_{\rm r}T^d \right)
 D^{cd\ [l]}_{\rho \sigma}(q) \right].
\label{JD}
\end{eqnarray}
Then the fermion self-energy part (\ref{JD}) is calculated as
\begin{eqnarray}
 J(k)
 =
 {g_{\rm r}^2 \over 16\pi^2}
 \Gamma \left( \epsilon l+ \epsilon \right)
 \left(- {4 \over 3} N_f T(R) g_{\rm r}^2 I \right)^l
 {\not \! k \over (-k^2)^{\epsilon (l+1)} }
 \left[-1+(1-\delta_{0l} \xi_{\rm r}) \right]
 T^c T^d  \delta^{cd} + O(m_{\rm r}^2).
\label{JDR}
\end{eqnarray}
When the gauge boson propagator
 $D^{cd \ [l]}_{\rho \sigma}(q)$ in (\ref{JD}) involves the
 fermion loop (i.e. $l \not= 0$),
 the term of $O(\epsilon^0)$ in (\ref{JDR}) vanishes,
 and then the vacuum polarization tensor
 $\Pi^{ab \ {\rm D}}_{\mu \nu }(p)$
 does not contribute to $Z_3$.
When $D^{cd \ [l]}_{\rho \sigma}(q)$ does not involve
 the fermion loop
 (i.e. $l=0$), (\ref{JDR}) is reduced to
\begin{eqnarray}
 J(k)
 =
 -\xi_{\rm r} {g_{\rm r}^2 \over 16\pi^2}
 \Gamma \left( \epsilon \right)
 {\not \! k \over (-k^2)^\epsilon }
 T^c T^d  \delta^{cd} + O(m_{\rm r}^2).
\label{JDRo}
\end{eqnarray}
Hence we obtain the contribution of $O(\epsilon^0)$ in
 diagram D with $l=0$,
 as follows
\begin{eqnarray}
 \Pi^{ab \ {\rm D}}_{\mu \nu}(p)
 =
 \xi_{\rm r} {g_{\rm r}^2 \over 8\pi^2}
 \left(- {4 \over 3} N_f T(R) g_{\rm r}^2 I \right)
 \Gamma (2\epsilon -1)
 {1 \over (-p^2)^{2\epsilon}}
 \left(g_{\mu \nu} p^2 - p_\mu p_\nu \right)
 {\rm Tr} \left(T^a T^b T^c T^d \right) \delta^{cd}.
\label{DR}
\end{eqnarray}
The contribution from diagram E is given by
\begin{eqnarray}
 \Pi^{ab \ {\rm E}}_{\mu \nu}(p)
 =
 -N_f \int {d^n k \over i(2\pi)^n } {\rm Tr}
 \left[ \Lambda^a_\mu (k){1 \over m_{\rm r} -(\not \! p + \not \! k) }
 \gamma_\nu \left(g_{\rm r} T^b \right)
 {1 \over m_{\rm r} -\not \! k} \right]
\label{E}
\end{eqnarray}
 with the vertex correction part
\begin{eqnarray}
 \Lambda^a_\mu (k)
 =
 2 \int {d^n q \over i(2\pi)^n }
 D^{cd \ [l]}_{\rho \sigma}(q-k)
 \gamma^\sigma \left( g_{\rm r}T^d \right)
 {1 \over m_{\rm r} -\not \! q} \gamma_\mu
 \left( g_{\rm r}T^a \right)
 {1 \over m_{\rm r} -(\not \! p + \not \! q) } \gamma^\rho
 \left( g_{\rm r}T^c \right) ,
\label{JE}
\end{eqnarray}
 where the factor $2$ arise from the overlapping divergence.
The vertex correction part (\ref{JE}) is calculated as
\begin{eqnarray}
 \Lambda^a_\mu (k)
 =
 {2g_{\rm r}^3 \over 16\pi^2}
 \left(- {4 \over 3} N_f T(R) g_{\rm r}^2 I \right)^l
 \Gamma \left(\epsilon l+\epsilon \right)
 {\gamma_\mu \over (-k^2)^{\epsilon (l+1)}}
 \left[1-(1-\delta_{0l} \xi_{\rm r} ) \right]
 T^d T^a T^c \delta^{cd} + O(m_{\rm r}^2).
\label{JER}
\end{eqnarray}
When the gauge boson propagator
  $D^{cd \ [l]}_{\rho \sigma}(q-k)$ in (\ref{JE}) involves
 the fermion loop (i.e. $l \not= 0$), the term of
 $O(\epsilon^0)$
 in (\ref{JER}) vanishes,
 and then $\Pi^{ab \ {\rm E}}_{\mu \nu}(p)$
 does not contribute to $Z_3$.
When $D^{cd \ [l]}_{\rho \sigma}(q-k)$ does not involve the fermion loop
 (i.e. $l=0$), (\ref{JER}) is reduced to
\begin{eqnarray}
 \Lambda^a_\mu (k)
 =
  \xi_{\rm r} {2g_{\rm r}^3 \over 16\pi^2}
 \Gamma (\epsilon )
 {\gamma_\mu \over (-k^2)^\epsilon }
 T^d T^a T^c \delta^{cd} + O(m_{\rm r}^2).
\label{JERo}
\end{eqnarray}
Hence we obtain the contribution of $O(\epsilon^0)$ in
 diagram E with
 $l=0$, as follows
\begin{eqnarray}
 \Pi^{ab \ {\rm E}}_{\mu \nu}(p)
 =
 -\xi_{\rm r}{2g_{\rm r}^2 \over 8\pi^2}
 \left(- {4 \over 3} N_f T(R) g_{\rm r}^2 I \right)
 \Gamma (2\epsilon -1)
 {1 \over (-p^2)^{2\epsilon}}
 \left(g_{\mu \nu} p^2 - p_\mu p_\nu \right)
 {\rm Tr} \left(T^d T^a T^c T^b \right) \delta^{cd}.
\label{ER}
\end{eqnarray}
The contribution from diagram F is given by
\begin{eqnarray}
 \Pi^{ab \ {\rm F}}_{\mu \nu }(p)
 =
 -{1 \over 2} \int {d^n q \over i(2\pi)^n }
 (ig_{\rm r} f^{bcc'})
 \Xi^{\nu \beta {\beta'}}(q,p)
 D^{cd \ [l]}_{\alpha \beta}(q)
 D^{{c'}{d'} \ [l']}_{\alpha' \beta'}(p-q)
 (ig_{\rm r} f^{add'})
 \Xi^{\mu \alpha {\alpha'}}(q,p)
\label{F}
\end{eqnarray}
 with
\begin{eqnarray}
 && \Xi^{\nu \beta {\beta'}}(q,p)
 =
 g^{\beta {\beta'}} (p-2q)^\nu
 + g^{{\beta'} \nu} (q-2p)^\beta
 + g^{\nu \beta} (p+q)^{\beta'}
\cr
&&
\Xi^{\mu \alpha {\alpha'}}(q,p)
 =
 g^{\alpha {\alpha'}} (p-2q)^\mu
 + g^{{\alpha'} \mu} (q-2p)^\alpha
 + g^{\mu \alpha} (p+q)^{\alpha'} ,
\label{JF}
\end{eqnarray}
 where $l$ and $l'$ are the numbers of one-fermion
 loops which
 are inserted into the each gauge boson propagator
 in Fig.\ref{FigOne} (F--H).
Then we obtain the contribution of $O(\epsilon^0)$ from diagram F,
\begin{eqnarray}
 \Pi^{ab \ {\rm F}}_{\mu \nu}(p)
 &=&
 {g_{\rm r}^2 \over 16\pi^2}
 \left(- {4 \over 3} N_f T(R) g_{\rm r}^2 I \right)^{l+l'}
 \Gamma \left( \epsilon l+\epsilon l' +\epsilon \right)
\cr
&&
 \times {1 \over (-p^2)^{\epsilon (l+l'+1)}}
  \left[ {25 \over 12}
 g_{\mu \nu} p^2 -{7 \over 3}p_\mu p_\nu
 -{1 \over 4} \xi_{\rm r} (\delta_{0l}+\delta_{0l'})
 \left(g_{\mu \nu} p^2 -p_\mu p_\nu \right) \right]
 C_2(G) \delta^{ab}.
\label{FR}
\end{eqnarray}
In (\ref{F}), when the gauge boson propagators
 $D^{cd \ [l]}_{\alpha \beta}(q)$ and
 $D^{{c'}{d'} \ [l']}_{\alpha' \beta' }(p-q)$ does not
 involve the fermion loop (i.e. $l=l'=0$),
 then (\ref{FR}) is equal to
 (\ref{BR}) which is the contribution of the diagram B.
Hereafter we suppress the mass of fermion in the propagators
 because
 the term involving $m_{\rm r}$ in $Z_3$ behave like $O(\epsilon)$
 or less, while we are retaining $O(\epsilon^0)$ terms,
 which are leading in $\epsilon$
 (see the first paragraph in this section).

The contribution from diagram G is given by
\begin{eqnarray}
 \Pi^{ab \ {\rm G}}_{\mu \nu}(p)
 = -{1 \over 2} \int {d^n q \over i(2\pi)^n }
 \int {d^n k \over i(2\pi)^n }
 (ig_{\rm r} f^{bcc'})
 \Xi^{\nu \beta {\beta'}}(q,p)
 D^{cd \ [l]}_{\alpha \beta }(q)
 D^{{c'}{d'} \ [l']}_{\alpha' \beta'}(p-q)
 {\cal F}^{ad{d'}}_{\mu \alpha \alpha'}(q,k,p)
\label{G}
\end{eqnarray}
 with the three boson vertex part
\begin{eqnarray}
 {\cal F}^{ad{d'}}_{\mu \alpha \alpha'}(q,k,p)
  &=&
 -N_f {\rm Tr}
 \left[
 {1 \over -(\not \! k + \not \! p)}
 ( g_{\rm r} T^a ) \gamma_\mu
 {1 \over -\not \! k}
 ( g_{\rm r} T^d ) \gamma_\alpha
 {1 \over -(\not \! k + \not \! q)}
 ( g_{\rm r} T^{d'} ) \gamma_{\alpha'}
 \right]
 \cr
 &&
 -N_f {\rm Tr}
 \left[
 {1 \over -(\not \! k - \not \! q)}
 ( g_{\rm r} T^d ) \gamma_\alpha
 {1 \over -\not \! k}
 ( g_{\rm r} T^a ) \gamma^\mu
 {1 \over -(\not \! k - \not \! p)}
 ( g_{\rm r} T^{d'} ) \gamma_{\alpha'}
 \right].
\label{JG}
\end{eqnarray}
We note that in (\ref{JG}) the fermion propagators
  $-(\not \! k + \not \! p)^{-1}$
 and $-(\not \! k - \not \! p)^{-1}$ are expressed as
\begin{eqnarray}
 {1 \over -(\not \! k + \not \! p)}
 ={1 \over -\not \! k}+{1 \over -\not \! k} \not \! p
 {1 \over -\not \! k}
 + {1 \over -\not \! k} \not \! p {1 \over -\not \! k} \not \! p
 {1 \over -(\not \! k + \not \! p) }
\label{JGEp}
\end{eqnarray}
 and 
\begin{eqnarray}
 {1 \over -(\not \! k - \not \! p)}
 ={1 \over -\not \! k}+{1 \over -\not \! k} (-\not \! p)
 {1 \over -\not \! k}
 + {1 \over -\not \! k} (-\not \! p) {1 \over -\not \! k}
 (-\not \! p)
 {1 \over -(\not \! k - \not \! p) },
\label{JGEm}
\end{eqnarray}
 respectively.
For overlapping divergence in the diagram G, we separate two parts
 as diagrams ${\rm G_f}$ and ${\rm G_m}$ (Fig.\ref{FigTwo}).
The subscript "f" indicates the contribution where the
 divergence occurs at
 the fermion loop subdiagram which is inserted to the
 three gauge boson vertex part (Fig.\ref{FigTwo} ${\rm G_f}$),
 while the subscript
 "m" indicates that where the divergence occurs at
 the boson-fermion-boson
 (mixed) loop subdiagram in the boson-fermion-fermion
 vertex part (Fig.\ref{FigTwo} ${\rm G_m}$).
When we retain only the leading divergence,
 the first and second terms in (\ref{JGEp}) and (\ref{JGEm})
 contribute to the diagram
 ${\rm G_f}$, and the third term contributes to the diagram
 ${\rm G_m}$.
The contribution from diagram ${\rm G_f}$ is given by
\begin{eqnarray}
 \Pi^{ab \ {\rm G_f}}_{\mu \nu }(p)
 =
 -{1 \over 2} \int {d^n q \over i(2\pi)^n }
 (ig_{\rm r} f^{bcc'})
 \Xi^{\nu \beta {\beta'}}(q,p)
 D^{cd \ [l]}_{\alpha \beta}(q)
 D^{{c'}{d'} \ [l']}_{\alpha' \beta'}(p-q)
 F^{ad{d'}}_{\mu \alpha \alpha'}(q,p)
\label{Gf}
\end{eqnarray}
 with the three gauge boson vertex part
\begin{eqnarray}
&& F^{ad{d'}}_{\mu \alpha \alpha'}(q,p)
\cr
&& =
 -g_{\rm r}^3 N_f
 \int {d^n k \over i(2\pi)^n }
 \Bigg\{
 {\rm Tr}(T^a T^d T^{d'})
 {\rm Tr}
 \left[ \left( 
 {1 \over -\not \! k }
 +{1 \over -\not \! k } \not \! p {1 \over -\not \! k }
 \right)
 \gamma_\mu {1 \over -\not \! k } \gamma_\alpha 
 {1 \over -(\not \! k + \not \! q)} \gamma_{\alpha'}
 \right]
\cr
&& \ \ \ \ \ \ \ \ \ \ \ \ \ \ \ \ \ \ \ 
  +
 {\rm Tr}(T^d T^a T^{d'})
 {\rm Tr} 
 \left[
 {1 \over -(\not \! k - \not \! q)} \gamma_\alpha
 {1 \over -\not \! k } \gamma_\mu
 \left(
 {1 \over -\not \! k}
 +{1 \over -\not \! k}(-\not \! p){1 \over -\not \! k}
 \right) 
 \gamma_{\alpha'}
 \right]
 \Bigg\}.
\label{JGf}
\end{eqnarray}
Then the three gauge boson vertex part
 $F^{ad{d'}}_{\mu \alpha \alpha'}(q,p)$ 
 is calculated as
\begin{eqnarray}
 F^{ad{d'}}_{\mu \alpha \alpha'}(q,p)
 = ig_{\rm r} f^{add'}
 \left(- {4 \over 3} N_f T(R) g_{\rm r}^2 I \right)
{1 \over (-q^2)^\epsilon}
 \left[ g_{\mu \alpha}(p+q)_{\alpha'}
 +g_{\mu \alpha'}(q-2p)_\alpha
 +g_{\alpha \alpha'}(p-2q)_\mu \right].
\label{JGfR}
\end{eqnarray}
Hence we obtain the contribution of $O(\epsilon^0)$ in
 diagram ${\rm G_f}$, as follows
\begin{eqnarray}
 \Pi^{ab \ {\rm G_f}}_{\mu \nu }(p)
 &=&
 -{g_{\rm r}^2 \over 16\pi^2}
 \left(-{4 \over 3} N_f T(R) g_{\rm r}^2 I \right)^{l+l'+1}
 \Gamma \left( \epsilon l+\epsilon l'+2\epsilon \right)
\cr
&&
 \times {1 \over (-p^2)^{\epsilon (l+l'+2)}}
  \left[
 {25 \over 12} g_{\mu \nu} p^2 -{7 \over 3}p_\mu p_\nu
 -{1 \over 4} \xi_{\rm r} (\delta_{0l}+\delta_{0l'})
 \left(g_{\mu \nu} p^2 -p_\mu p_\nu \right)
 \right] C_2(G) \delta^{ab}.
\label{GfR}
\end{eqnarray}
The contribution of diagram ${\rm G_m}$ is given by
\begin{eqnarray}
 \Pi^{ab \ {\rm G_m}}_{\mu \nu}(p)
  &=&
 -{1 \over 2} \int {d^n k \over i(2\pi)^n }
 (ig_{\rm r}f^{bcc'})g_{\rm r}^3 N_f
\cr
&&
 \times
\Bigg\{ 
 {\rm Tr}(T^a T^d T^{d'})
 {\rm Tr} 
 \left[
 {1 \over -\not \! k} \not \! p 
 {1 \over -\not \! k} \not \! p
 {1 \over -(\not \! k + \not \! p)} \gamma_\mu 
 {1 \over -\not \! k} 
 K^{cdc'd'}_{\nu\ +}(k,p)
 \right]
\cr
&&
 +
 {\rm Tr}(T^d T^a T^{d'})
 {\rm Tr} 
 \left[
 {1 \over -\not \! k } \gamma_\mu 
 {1 \over -\not \! k } \not \! p
 {1 \over -\not \! k } \not \! p
 {1 \over -(\not \! k - \not \! p)}
 K^{cdc'd'}_{\nu\ -}(k,p)
 \right]
 \Bigg\}.
\label{Gm}
\end{eqnarray}
 with the boson-fermion-fermion vertex parts
\begin{eqnarray}
 K^{cdc'd'}_{\nu\ +} (k,p)
 =
 \int {d^n q \over i(2\pi)^n }
 \gamma_\alpha \Xi_{\nu \beta {\beta'}}(q,p)
 D^{\alpha \beta \ cd \ [l]}(q)
 D^{\alpha' \beta' \ {c'}{d'} \ [l']}(p-q)
 {1 \over -(\not \! k + \not \! q)}
 \gamma_{\alpha'}
\label{Kp}
\end{eqnarray}
 and
\begin{eqnarray}
K^{cdc'd'}_{\nu\ -}(k,p)
 =
 \int {d^n q \over i(2\pi)^n }
 \gamma_{\alpha'}
 \Xi_{\nu \beta {\beta'}}(q,p)
 D^{\alpha \beta \ cd \ [l]}(q)
 D^{\alpha' \beta' \ {c'}{d'} \ [l']}(p-q)
 {1 \over -(\not \! k - \not \! q)}
 \gamma_\alpha .
\label{Km}
\end{eqnarray}
Then the boson-fermion-fermion vertex parts
 $K^{cdc'd'}_{\nu\ +}(k,p)$
 and $K^{cdc'd'}_{\nu\ -}(k,p)$
 are calculated as
\begin{eqnarray}
 K^{cdc'd'}_{\nu\ +} (k,p)
 &=& {1 \over 16\pi^2}
 \left(-{4 \over 3} N_f T(R) g_{\rm r}^2 I \right)^{l+l'}
 \Gamma(\epsilon l+ \epsilon l'+ \epsilon)
 {\gamma_\nu \over (-k^2)^{\epsilon (l+l'+1)}}
 \left[ -{3 \over 2}-{3 \over 4}\xi_{\rm r}(\delta_{0l}+\delta_{0l'})
  \right] \delta^{cd} \delta^{c'd'}
\cr
&&
 + O(m_{\rm r}^2)
\label{KmpR}
\end{eqnarray}
 and $K^{cdc'd'}_{\nu\ -} (k,p) = -K^{cdc'd'}_{\nu\ +} (k,p)$.
Hence we obtain the contribution of $O(\epsilon^0)$ in diagram
 ${\rm G_m}$, as follows
\begin{eqnarray}
 \Pi^{ab \ {\rm G_m}}_{\mu \nu }(p)
 &=&
 {g_{\rm r}^2 \over 16\pi^2 \Gamma ( \epsilon )}
 \left(- {4 \over 3} N_f T(R) g_{\rm r}^2 I \right)^{l+l'+1}
 \Gamma \left( \epsilon l+\epsilon l'+\epsilon \right)
 \Gamma \left( \epsilon l+\epsilon l'+2\epsilon \right)
\cr
&&
 \times {1 \over (-p^2)^{\epsilon (l+l'+2)}}
  \left[
 {3 \over 4}+{3 \over 8} \xi_{\rm r} (\delta_{0l}+\delta_{0l'}) \right]
 \left(g_{\mu \nu} p^2 -p_\mu p_\nu \right)
 C_2(G) \delta^{ab}.
\label{GmR}
\end{eqnarray}
For the diagrams ${\rm H_f}$ and ${\rm H_m}$, we can discuss the
 contributions to $Z_3$ in a similar manner to the diagrams
 ${\rm G_f}$ and ${\rm G_m}$.
The contributions from diagrams ${\rm H_f}$ and ${\rm H_m}$ are
 calculated as
\begin{eqnarray}
 \Pi^{ab \ {\rm H_f}}_{\mu \nu}(p)
 &=&
 {g_{\rm r}^2 \over 16\pi^2}
 \left(- {4 \over 3} N_f T(R) g_{\rm r}^2 I \right)^{l+l'+2}
 \Gamma \left( \epsilon l+ \epsilon l' +3\epsilon \right)
\cr
&&
 \times {1 \over (-p^2)^{\epsilon (l+l'+3)}}
 \left[
 {25 \over 12}g_{\mu \nu} p^2 -{7 \over 3}p_\mu p_\nu
 -{1 \over 4} \xi_{\rm r} (\delta_{0l}+\delta_{0l'})
 \left(g_{\mu \nu} p^2 -p_\mu p_\nu \right)
 \right]
 C_2(G) \delta^{ab}
\label{HfR}
\end{eqnarray}
 and
\begin{eqnarray}
 \Pi^{ab \ {\rm H_m}}_{\mu \nu}(p)
 &=&
 {g_{\rm r}^2 \over 16\pi^2 \Gamma (\epsilon )}
 \left(- {4 \over 3} N_f T(R) g_{\rm r}^2 I \right)^{l+l'+2}
 \Gamma \left(\epsilon l+\epsilon l'+2\epsilon \right)
 \Gamma \left(\epsilon l+\epsilon l'+3\epsilon \right)
\cr
&&
 \times {1 \over (-p^2)^{\epsilon (l+l'+3)}}
  \left[
 -{3 \over 4} - {3 \over 8} \xi_{\rm r} (\delta_{0l}+\delta_{0l'})
 \right]
 \left(g_{\mu \nu} p^2 -p_\mu p_\nu \right)
 C_2(G) \delta^{ab},
\label{HmR}
\end{eqnarray}
 respectively.

Next we discuss the total contributions from the diagrams in Fig.1.
Hereafter we note {\it the number of one-fermion loops per the diagram}
 with the notation $\widetilde l$.
Then the numbers $\widetilde l$ to the diagrams F, G, and H are given
 by $l+l'$, $l+l' +1$, and $l+l' +2$, respectively.
For $\widetilde l =0$, the diagrams G and H are absent while
 the diagram F reduces to the diagram B.

First let us consider the contributions of
 {\it gauge independent part} to $Z_3$.
The {\it gauge independent part} in the vacuum
 polarization tensor is noted
 with the subscript "0" on the tensor.
In the total contribution from diagrams D and E,
 the {\it gauge independent part} does not contribute to $Z_3$
 from (\ref{DR}) and (\ref{ER}).
For $\widetilde l \geq 1$, {\it multiplicities} of diagrams
 F, ${\rm G_f}$, and ${\rm H_f}$
 are $\widetilde l+1$, $2\widetilde l$,
 and $\widetilde l-1$, respectively, where
 the {\it multiplicities} is the numbers of diagrams with the
 same contribution to $Z_3$ for a number $\widetilde l$.
In the total contribution from diagrams F, ${\rm G_f}$,
 and ${\rm H_f}$, when $\widetilde l \geq 1$,
 the {\it gauge independent parts} cancel each other, as follows
\begin{eqnarray}
 && (\widetilde l +1) \Pi^{ab \ {\rm F}}_{\mu \nu\ 0}(p)
 + 2 \widetilde l \Pi^{ab \ {\rm G_f}}_{\mu \nu\ 0}(p)
 + (\widetilde l -1) \Pi^{ab \ {\rm H_f}}_{\mu \nu\ 0}(p)
 \cr
 &&= {g_{\rm r}^2 \over 16\pi^2}
 \left[(\widetilde l+ 1)-2 \widetilde l +(\widetilde l -1) \right]
 \left(- {4 \over 3} N_f T(R) g_{\rm r}^2 I \right)^{\widetilde l}
 {\Gamma (\epsilon \widetilde l +\epsilon)
 \over (-p^2)^{\epsilon (\widetilde l +1)}}
 \left({25 \over 12} g_{\mu \nu} p^2 -{7 \over 3}p_\mu p_\nu \right)
 C_2(G) \delta^{ab}
 \cr
 &&= 0
\label{FGfHf}
\end{eqnarray}
 from (\ref{FR}), (\ref{GfR}), and (\ref{HfR}).
For $\widetilde l \geq 2$, the multiplicities of diagrams
 ${\rm G_m}$ and ${\rm H_m}$ are $2\widetilde l$ and
 $2(\widetilde l-1)$, respectively.
In the total contribution from diagrams ${\rm G_m}$
 and ${\rm H_m}$, when $\widetilde l \geq 2$,
 the {\it gauge independent part} is given by
\begin{eqnarray}
 &&  2 \widetilde l \Pi^{ab \ {\rm G_m}}_{\mu \nu\ 0}(p)
 + 2(\widetilde l -1) \Pi^{ab \ {\rm H_m}}_{\mu \nu\ 0}(p)
\cr
 &&=  { g_{\rm r}^2 \over 16\pi^2 \Gamma (\epsilon)}
 \left[ 2 \widetilde l - 2(\widetilde l -1) \right]
 \left(- {4 \over 3} N_f T(R) g_{\rm r}^2 I \right)^{\widetilde l}
 {\Gamma (\epsilon \widetilde l)\
  \Gamma (\epsilon \widetilde l + \epsilon)
  \over (-p^2)^{\epsilon (\widetilde l +1)}}
 \left({3 \over 4} g_{\mu \nu} p^2 - {3 \over 4} p_\mu p_\nu \right)
 C_2(G) \delta^{ab}
\cr
 && \approx {3 \over 2}
 {g_{\rm r}^2 I \over \widetilde l ( \widetilde l + 1 )}
 \left(- {4 \over 3} N_f T(R) g_{\rm r}^2 I \right)^{\widetilde l}
 \left(g_{\mu \nu} p^2 -p_\mu p_\nu \right)
 C_2(G) \delta^{ab}
\label{GmHm}
\end{eqnarray}
 from (\ref{GmR}) and (\ref{HmR}).
Hence the diagrams ${\rm G_m}$ and ${\rm H_m}$ with
 many one-fermion loops (i.e. $\widetilde l \geq 2$)
 contribute to $Z_3$ in total.

Now we consider the contributions of {\it gauge dependent part} to $Z_3$.
The {\it gauge dependent part} in the vacuum
 polarization tensor is noted
 with the subscript "$\xi$" on the tensor.
For $\widetilde l = 1$, the multiplicities of diagrams
 D and E are 2 and 1, respectively.
From (\ref{DR}) and (\ref{ER}), the total contribution from
 diagrams D and E is given by
\begin{eqnarray}
  2 \Pi^{ab \ {\rm D}}_{\mu \nu\ \xi}(p)
  + \Pi^{ab \ {\rm E}}_{\mu \nu\ \xi}(p)
 &=&
 \xi_{\rm r}{g_{\rm r}^2 \over 16\pi^2}
 \left(- {4 \over 3} N_f T(R) g_{\rm r}^2 I \right)
 {\Gamma (2\epsilon -1) \over (-p^2)^{2\epsilon} }
 \left(g_{\mu \nu} p^2 - p_\mu p_\nu \right)
 C_2(G) \delta^{ab}
\cr
 & \approx &
 -\xi_{\rm r}{ g_{\rm r}^2 I \over 2}
 \left(- {4 \over 3} N_f T(R) g_{\rm r}^2 I \right)
 \left(g_{\mu \nu} p^2 - p_\mu p_\nu \right) C_2(G) \delta^{ab} ,
\label{DER}
\end{eqnarray}
 where we have used relation
\begin{eqnarray}
 {\rm Tr} \left( T^a T^b T^c T^c \right)
 -{\rm Tr} \left( T^c T^a T^c T^b \right)
 =
 {1 \over 4} C_2(G) \delta^{ab}.
\label{TR}
\end{eqnarray}
Notice that the diagrams D and E with the many one-fermion loops
 does not contribute to $Z_3$.
To the {\it gauge dependent part},
 the diagrams F, G, and H have contributions
 only when one-fermion loop is not inserted on the
 gauge boson propagator, and then the factor
 $\delta_{0l}+\delta_{0l'}$ give $2$
 in (\ref{FR}), (\ref{GfR}), (\ref{GmR}), (\ref{HfR}), and (\ref{HmR}).
For the {\it gauge dependent part} with $\widetilde l =1$,
 we obtain the total contribution from diagrams D, E, F, ${\rm G_f}$,
 and ${\rm G_m}$ to $Z_3$,
\begin{eqnarray}
 &&  2\Pi^{ab \ {\rm D}}_{\mu \nu\ \xi}(p)
    + \Pi^{ab \ {\rm E}}_{\mu \nu\ \xi}(p)
    + \Pi^{ab \ {\rm F}}_{\mu \nu\ \xi}(p)
    +2\Pi^{ab \ {\rm G_f}}_{\mu \nu\ \xi}(p)
    +2\Pi^{ab \ {\rm G_m}}_{\mu \nu\ \xi}(p)
\cr
 && \approx
 \xi_{\rm r} g_{\rm r}^2 I
 \left(-{1 \over 2}-{1 \over 4}+{1 \over 2}+{3 \over 4} \right)
 \left(- {4 \over 3} N_f T(R) g_{\rm r}^2 I \right)
 \left(g_{\mu \nu}p^2 -p_\mu p_\nu \right)
 C_2(G) \delta^{ab}
\cr
 &&= \xi_{\rm r}{g_{\rm r}^2 I \over 2}
 \left(- {4 \over 3} N_f T(R) g_{\rm r}^2 I \right)
 \left(g_{\mu \nu}p^2 -p_\mu p_\nu \right)
 C_2(G) \delta^{ab}
\label{DEFGfGmK}
\end{eqnarray}
 from (\ref{FR}), (\ref{GfR}), (\ref{GmR}), and (\ref{DER}).
For $\widetilde l \geq 2$, the multiplicities of diagrams
 F, ${\rm G_f}$, and ${\rm H_f}$ are 1, 2, 1, respectively.
In the total contribution from diagrams F, ${\rm G_f}$,
 and ${\rm H_f}$, when $\widetilde l \geq 2$,
 the {\it gauge dependent parts} cancel each other, as follows
\begin{eqnarray}
 && \Pi^{ab \ {\rm F}}_{\mu \nu\ \xi}(p)
    + 2\Pi^{ab \ {\rm G_f}}_{\mu \nu\ \xi}(p)
    + \Pi^{ab \ {\rm H_f}}_{\mu \nu\ \xi}(p)
\cr
 &&=
 \xi_{\rm r}{(1-2+1)g_{\rm r}^2 \over 16\pi^2}
 \left(- {4 \over 3} N_f T(R) g_{\rm r}^2 I \right)^{\widetilde l}
 { \Gamma (\epsilon \widetilde l +\epsilon) \over
 (-p^2)^{\epsilon (\widetilde l +1)}}
  \left(-{2 \over 4} g_{\mu \nu} p^2 + {2 \over 4} p_\mu p_\nu \right)
 C_2(G) \delta^{ab}
\cr
 &&=0
\label{FGfHfK}
\end{eqnarray}
 from (\ref{FR}), (\ref{GfR}), and (\ref{HfR}).
For $\widetilde l \geq 2$, the multiplicities of diagrams
 ${\rm G_m}$ and ${\rm H_m}$ are 2 and 2, respectively.
In the total contribution from diagrams ${\rm G_m}$ and
 ${\rm H_m}$, when $\widetilde l \geq 2$,
 the {\it gauge dependent parts} cancel each other, as follows
\begin{eqnarray}
&& 2\Pi^{ab \ {\rm G_m}}_{\mu \nu\ \xi}(p)
    + 2\Pi^{ab \ {\rm H_m}}_{\mu \nu\ \xi}(p)
\cr
 &&=
 \xi_{\rm r}{(2-2)g_{\rm r}^2 \over 16\pi^2 \Gamma (\epsilon)}
 \left(- {4 \over 3} N_f T(R) g_{\rm r}^2 I \right)^{\widetilde l}
 { \Gamma (\epsilon \widetilde l)\
 \Gamma (\epsilon \widetilde l +\epsilon) \over
 (-p^2)^{\epsilon (\widetilde l +1)}}
   \left(-{6 \over 8}g_{\mu \nu} p^2 + {6 \over 8} p_\mu p_\nu \right)
 C_2(G) \delta^{ab}
\cr
 &&=0
\label{GmHmK}
\end{eqnarray}
 from (\ref{GmR}) and (\ref{HmR}).

\subsection{Renormalization of subdiagram divergences}

We discuss the renormalization to the subdiagram divergences.
The divergent parts in diagrams are each one-fermion loop,
 the fermion self-energy part in D, the fermion-boson
 vertex part in E, the three-boson vertex part in
 ${\rm G_f}$ and ${\rm H_f}$, and
 the boson-fermion-fermion vertex part in
 ${\rm G_m}$ and ${\rm H_m}$.
Hereafter we consider the minimal subtraction scheme.

For the divergence from one-fermion loop diagram,
 the counter term is given by
\begin{eqnarray}
 \Pi^{ab \ {\rm A}}_{\mu \nu\ {\rm c.t.}}(p)
 = -{4 \over 3}{1 \over (4\pi)^2} N_f g_{\rm r}^2 \Gamma(\epsilon)
 \left(-g_{\mu \nu} p^2 + p_\mu p_\nu \right) {\rm Tr}(T^a T^b)
\label{ARc}
\end{eqnarray}
 from (\ref{A}).
Then the gauge boson propagator (\ref{GPN})
 with many one-fermion loops
 is replaced by renormalized one, as follows
\begin{eqnarray}
 D^{ab\ [l]}_{\mu \nu\ {\rm ren}}(p)
 &=&
 \left(-{4 \over 3}N_f T(R) g_{\rm r}^2 I \right)^l
  \left[ {1 \over (-p^2)^\epsilon }- 1 \right]^l
  {1 \over (-p^2)} \left[-g_{\mu \nu}+(1-\xi_{\rm r} \delta_{0 l})
  {p_\mu p_\nu \over p^2} \right] \delta^{ab}
 \cr
 &=&
 \left(-{4 \over 3}N_f T(R) g_{\rm r}^2 I \right)^l
 \sum_{n=0}^l {l! \over n!(l-n)!}
  {(-1)^{l-n} \over (-p^2)^{1+\epsilon n}}
  \left[-g_{\mu \nu}+(1-\xi_{\rm r} \delta_{0 l})
 {p_\mu p_\nu \over p^2} \right] \delta^{ab} ,
\label{GPREN}
\end{eqnarray}
 where $n$ is the number of one-fermion loops in the gauge boson
 propagator, and $l-n$ is the number of
 counter terms in the gauge boson propagator
 for the divergences from one-fermion loops.

To the divergences of fermion self-energy part (\ref{JDRo}) and
 vertex correction part (\ref{JERo}),
 the counter terms are given by
\begin{eqnarray}
  J_{\rm c.t.}(k)
 =
 \xi_{\rm r} {g_{\rm r}^2 \over 16\pi^2}
 \Gamma \left( \epsilon \right)
 \not \! k
 T^c T^d  \delta^{cd}
\label{JDRoc}
\end{eqnarray}
 and
\begin{eqnarray}
 \Lambda^a_{\mu\ {\rm c.t.}}
 =
  - \xi_{\rm r} {2g_{\rm r}^3 \over 16\pi^2}
 \Gamma (\epsilon )\gamma_\mu T^d T^a T^c \delta^{cd} ,
\label{JERoc}
\end{eqnarray}
 respectively.
Then (\ref{DR}), (\ref{ER}), and (\ref{DER})
 are replaced by
\begin{eqnarray}
 \Pi^{ab \ {\rm D}}_{\mu \nu\ {\rm ren}}(p)
 =
 \xi_{\rm r} {g_{\rm r}^2 \over 8\pi^2}
 \left(- {4 \over 3} N_f T(R) g_{\rm r}^2 I \right)
 \left[ { \Gamma (2\epsilon - 1) \over (-p^2)^{2\epsilon} }
 - { \Gamma (\epsilon - 1) \over (-p^2)^{\epsilon} } \right]
 \left(g_{\mu \nu} p^2 - p_\mu p_\nu \right)
 {\rm Tr} \left(T^a T^b T^c T^d \right) \delta^{cd},
\label{DRr}
\end{eqnarray}
\begin{eqnarray}
 \Pi^{ab \ {\rm E}}_{\mu \nu\ {\rm ren}}(p)
 =
 -\xi_{\rm r}{2g_{\rm r}^2 \over 8\pi^2}
 \left(- {4 \over 3} N_f T(R) g_{\rm r}^2 I \right)
 \left[ { \Gamma (2\epsilon -1) \over (-p^2)^{2\epsilon} }
 - { \Gamma (\epsilon -1) \over (-p^2)^{\epsilon} } \right]
 \left(g_{\mu \nu} p^2 - p_\mu p_\nu \right)
 {\rm Tr} \left(T^d T^a T^c T^b \right) \delta^{cd},
\label{ERr}
\end{eqnarray}
 and
\begin{eqnarray}
 && 2 \Pi^{ab \ {\rm D}}_{\mu \nu\ \xi \ {\rm ren}}(p)
    + \Pi^{ab \ {\rm E}}_{\mu \nu\ \xi \ {\rm ren}}(p)
\cr
&&
 =
 \xi_{\rm r}{g_{\rm r}^2 \over 16\pi^2}
 \left(- {4 \over 3} N_f T(R) g_{\rm r}^2 I \right)
 \left[
 {\Gamma (2\epsilon -1) \over (-p^2)^{2\epsilon} }
 -{\Gamma (\epsilon -1) \over (-p^2)^{\epsilon} }
 \right]
 \left(g_{\mu \nu} p^2 - p_\mu p_\nu \right)
 C_2(G) \delta^{ab}
\cr
&&
 \approx 
 \xi_{\rm r}{ g_{\rm r}^2 I \over 2}
 \left(- {4 \over 3} N_f T(R) g_{\rm r}^2 I \right)
 \left(g_{\mu \nu} p^2 - p_\mu p_\nu \right) C_2(G) \delta^{ab} ,
\label{DERr}
\end{eqnarray}
 respectively.

We take into account the counter terms for the divergence
 from many one-fermion loops in the diagrams F, G, and H. 
Then (\ref{FR}) is replaced by
\begin{eqnarray}
 \Pi^{ab \ {\rm F}}_{\mu \nu\ {\rm ren}}(p)
 &=&
 {g_{\rm r}^2 \over 16\pi^2}
 \left(- {4 \over 3} N_f T(R) g_{\rm r}^2 I \right)^{\widetilde l}
 \sum_{{\widetilde n}=0}^{\widetilde l}
 {{\widetilde l}! \over {\widetilde n}!({\widetilde l}-{\widetilde n})!}
 \Gamma \left( \epsilon {\widetilde n} +\epsilon \right)
 {(-1)^{{\widetilde l}-{\widetilde n}} \over
 (-p^2)^{\epsilon ({\widetilde n}+1)}}
\cr
&&
 \times
  \left[ {25 \over 12}
 g_{\mu \nu} p^2 -{7 \over 3}p_\mu p_\nu
 -{1 \over 4} \xi_{\rm r} (\delta_{0l}+\delta_{0l'})
 \left(g_{\mu \nu} p^2 -p_\mu p_\nu \right) \right]
 C_2(G) \delta^{ab}
\label{FRr}
\end{eqnarray}
 with $\widetilde l =l+l'$ and $\widetilde n =n+n'$,
 where $n$ and $n'$ are numbers of one fermion loops in each gauge boson
 propagator, and $l-n$ and $l'-n'$ are numbers of counter terms for the divergence
 from one-fermion loops in each gauge boson propagator.
To the divergence from three boson vertex part (\ref{JGfR}),
 the counter term is given by
\begin{eqnarray}
 F^{ad{d'}}_{\mu \alpha \alpha'\ {\rm c.t.}}(q,p)
 = -ig_{\rm r} f^{add'}
 \left(- {4 \over 3} N_f T(R) g_{\rm r}^2 I \right)
 \left[ g_{\mu \alpha}(p+q)_{\alpha'}
 +g_{\mu \alpha'}(q-2p)_\alpha
 +g_{\alpha \alpha'}(p-2q)_\mu \right].
\label{JGfRc}
\end{eqnarray}
Then we replace (\ref{GfR}) and (\ref{GmR}) by
\begin{eqnarray}
 \Pi^{ab \ {\rm G_f}}_{\mu \nu\ {\rm ren}}(p)
 &=&
 {g_{\rm r}^2 \over 16\pi^2}
 \left(- {4 \over 3} N_f T(R) g_{\rm r}^2 I \right)^{\widetilde l}
 \sum_{\widetilde n =0}^{\widetilde l -1}
 {(\widetilde l -1)! \over {\widetilde n}!({\widetilde l}-{\widetilde n}-1)!}
 \left[ {1 \over (-p^2)^{\epsilon}} \Gamma ( \epsilon \widetilde n
 +2\epsilon )
  - \Gamma (\epsilon \widetilde n + \epsilon )
 \right]
\cr
&&
 \times
 {(-1)^{\widetilde l - \widetilde n } \over (-p^2)^{\epsilon (\widetilde n+1)}}
  \left[ {25 \over 12}g_{\mu \nu} p^2 -{7 \over 3}p_\mu p_\nu
 -{1 \over 4} \xi_{\rm r} (\delta_{0l}+\delta_{0l'})
 \left(g_{\mu \nu} p^2 -p_\mu p_\nu \right) \right]
 C_2(G) \delta^{ab}
\label{GfRr}
\end{eqnarray}
 and 
\begin{eqnarray}
 \Pi^{ab \ {\rm G_m}}_{\mu \nu\ {\rm ren}}(p)
 &=&
 {g_{\rm r}^2 \over 16\pi^2}
 \left(- {4 \over 3} N_f T(R) g_{\rm r}^2 I \right)^{\widetilde l}
 \sum_{{\widetilde n}=0}^{\widetilde l -1}
 {(\widetilde l -1)! \over \widetilde n !(\widetilde l -\widetilde n -1)!}
 \Gamma ( \epsilon \widetilde n + \epsilon )
 \left[ {1 \over (-p^2)^{\epsilon \widetilde n}}
 {\Gamma ( \epsilon \widetilde n + 2\epsilon ) \over \Gamma (\epsilon) } - 1
 \right]
\cr
&&
 \times
 {(-1)^{\widetilde l - \widetilde n} \over (-p^2)^\epsilon }
  \left[
 -{3 \over 4}-{3 \over 8} \xi_{\rm r} (\delta_{0l}+\delta_{0l'}) \right]
 \left(g_{\mu \nu} p^2 -p_\mu p_\nu \right)
 C_2(G) \delta^{ab},
\label{GmRr}
\end{eqnarray}
 respectively,
 where $\widetilde l =l+l'+1$ and $\widetilde n =n+n'$.
In the similar manner, (\ref{HfR}) and (\ref{HmR})
 are replaced by
\begin{eqnarray}
 \Pi^{ab \ {\rm H_f}}_{\mu \nu\ {\rm ren}}(p)
 = - \Pi^{ab \ {\rm G_f}}_{\mu \nu\ {\rm ren}}(p)
\label{HfRr}
\end{eqnarray}
\begin{eqnarray}
 \Pi^{ab \ {\rm H_m}}_{\mu \nu\ {\rm ren}}(p)
 = - \Pi^{ab \ {\rm G_m}}_{\mu \nu\ {\rm ren}}(p)
\label{HmRr}
\end{eqnarray}
 with $\widetilde l =l+l'+2$ and $\widetilde n =n+n'$.

The total contribution (\ref{GmHm})
 is replaced by
\begin{eqnarray}
 &&  2 \widetilde l \Pi^{ab \ {\rm G_m}}_{\mu \nu\ 0\ {\rm ren}}(p)
 + 2(\widetilde l -1) \Pi^{ab \ {\rm H_m}}_{\mu \nu\ 0\ {\rm ren}}(p)
\cr
 &&
 \approx
 \left[2\widetilde l -2(\widetilde l -1) \right]
 {g_{\rm r}^2 \over 16\pi^2 \epsilon}
 \left(- {4 \over 3} N_f T(R) g_{\rm r}^2 I \right)^{\widetilde l}
 \sum_{{\widetilde n}=0}^{\widetilde l -1}
 {(\widetilde l -1)! \over \widetilde n !(\widetilde l -\widetilde n -1)!}
  { 1 \over (\widetilde n + 1)}
 \left( {1 \over \widetilde n +2} -1 \right)
 (-1)^{\widetilde l - \widetilde n}
\cr
&&
 \times
 \left(-{3 \over 4}g_{\mu \nu} p^2 +{3 \over 4}p_\mu p_\nu \right)
 C_2(G) \delta^{ab}
\cr
&&
 =
 {3 \over 2}
 {g_{\rm r}^2 I \over \widetilde l ( \widetilde l + 1 )}
 \left( {4 \over 3} N_f T(R) g_{\rm r}^2 I \right)^{\widetilde l}
 \left(g_{\mu \nu} p^2 -p_\mu p_\nu \right)
 C_2(G) \delta^{ab}
\label{GmHmr}
\end{eqnarray}
 from (\ref{GmRr}) and (\ref{HmRr}). 
The total contribution (\ref{DEFGfGmK}) is replaced by
\begin{eqnarray}
 &&  2\Pi^{ab \ {\rm D}}_{\mu \nu\ \xi \ {\rm ren}}(p)
    + \Pi^{ab \ {\rm E}}_{\mu \nu\ \xi \ {\rm ren}}(p)
    + \Pi^{ab \ {\rm F}}_{\mu \nu\ \xi \ {\rm ren}}(p)
    +2\Pi^{ab \ {\rm G_f}}_{\mu \nu\ \xi \ {\rm ren}}(p)
    +2\Pi^{ab \ {\rm G_m}}_{\mu \nu\ \xi \ {\rm ren}}(p)
\cr
 && \approx
  \xi_{\rm r}{g_{\rm r}^2 I \over 2}
 \left( {4 \over 3} N_f T(R) g_{\rm r}^2 I \right)
 \left(g_{\mu \nu}p^2 -p_\mu p_\nu \right)
 C_2(G) \delta^{ab}
\label{DEFGfGmKr}
\end{eqnarray}
 from (\ref{DERr}), (\ref{FRr}), (\ref{GfRr}), and (\ref{GmRr}).

Hence we obtain the total contribution from all diagrams in Fig.\ref{FigOne},
 as follows
\begin{eqnarray}
 \Pi^{ab}_{\mu \nu \ {\rm div}}
 &=&
 \left[
 -{4 \over 3} g_{\rm r}^2 I N_f T(R)
 + {2 \over 3} \xi_{\rm r} (g_{\rm r}^2 I)^2 N_f T(R) C_2(G)
 + \left( {13 \over 6}-{1 \over 2}
 \xi_{\rm r} \right) g_{\rm r}^2 I C_2(G)
 \right]
 (g_{\mu \nu} p^2 - p_\mu p_\nu ) \delta^{ab}
 \cr
 &&
 + \sum_{\widetilde l =1}^\infty
 {2 \over \widetilde l (\widetilde l + 1)}
 \left({4 \over 3} \right)^{\widetilde l -1}
 (g_{\rm r}^2 I)^{\widetilde l +1} \left[N_f T(R) \right]^{\widetilde l}
 (g_{\mu \nu} p^2 - p_\mu p_\nu ) C_2(G) \delta^{ab}
\label{AMOUNT}
\end{eqnarray}
 from (\ref{GmHmr}) and (\ref{DEFGfGmKr}). 
Then the renormalization constant $Z_3$ is
 given by
\begin{eqnarray}
 Z_3
 &=&
 1-{4 \over 3}g_{\rm r}^2 I N_f T(R)
   + \sum_{\widetilde l =1}^\infty
 {2 \over \widetilde l (\widetilde l + 1)}
 \left({4 \over 3} \right)^{\widetilde l -1}
 (g_{\rm r}^2 I)^{\widetilde l +1}
 \left[N_f T(R) \right]^{\widetilde l} C_2(G)
 \cr
 &&
 + {2 \over 3}\xi_{\rm r} (g_{\rm r}^2 I)^2 N_f T(R) C_2(G)
 + g_{\rm r}^2 I \left( {13 \over 6}-{1 \over 2}\xi_{\rm r} \right)
  C_2(G).
\label{ZNEXT}
\end{eqnarray}
Therefore we obtain the renormalization
 constant $Z_3$ to non-abelian gauge field
 at the leading and the next-to-leading order
 in $N_f^{-1}$, as follows
\begin{eqnarray}
 Z_3
 &=&
 1-{4 \over 3}g_{\rm r}^2 I N_f T(R)
 +{11 \over 3} g_{\rm r}^2 I C_2(G)
 -{1 \over 2} \xi_{\rm r} g_{\rm r}^2 I C_2(G)
 \left[ 1- {4 \over 3} g_{\rm r}^2 I N_f T(R) \right]
 \cr
 &&
 +{3 \over 2} g_{\rm r}^2 I C_2(G)
  \left[ {4 \over 2 g_{\rm r}^2 I N_f T(R)} - 1 \right]
 {\rm log}_e \left[ 1- {4 \over 3} g_{\rm r}^2 I N_f T(R) \right]
\label{ZRESULT}
\end{eqnarray}
 with $I \approx 1/(16\pi^2 \epsilon)$.
In (\ref{ZRESULT}), the third, foruth, and fifth terms
 arise from the
 next-to-leading contribution in the $1/N_f$ expansion.

\section{ SOLVING THE COMPOSITENESS CONDITION }

We have discussed so far the renormalization constant $Z_3$ in the general
 non-abelian gauge field theory (\ref{LG}) at the next-to-leading
 order in $1/N_f$.
The information of composite gauge field theory (\ref{LA}) can be obtained
 from the gauge field theory with $Z_3=0$.
Now we use the compositeness condition on
 (\ref{ZRESULT}) and solve it for $g_{\rm r}$.
The equation (\ref{g0L}) shows that
\begin{eqnarray}
 g_{\rm r}^2={3 \over 4N_f T(R)I} + O \left({1 \over N_f^2} \right).
\label{CG}
\end{eqnarray}
Substituting (\ref{CG}) to the equation $Z_3=0$ with (\ref{ZRESULT}),
 we see that the logarithmic term and the gauge dependent term in
 (\ref{ZRESULT}) are the order of $1/N_f^2$, and then these terms are
 negligible in the present approximation.
Hence $Z_3=0$ reduces to the simple equation
\begin{eqnarray}
 1-{4 \over 3}g_{\rm r}^2 I N_f T(R)
 +{11 \over 3} g_{\rm r}^2 I C_2(G) =0.
\label{ZCS}
\end{eqnarray}
Therefore we obtain
\begin{eqnarray}
 g_{\rm r}^2 = {3 \over \left[4N_f T(R) -11C_2(G) \right] I}
\label{COUPLING}
\end{eqnarray}
 at the next-to-leading order in $1/N_f$ expansion.
In (\ref{COUPLING}), if the following relation holds,
\begin{eqnarray}
  N_f  > {11C_2(G) \over 4T(R)} ,
\label{COND}
\end{eqnarray}
 $g_{\rm r}^2$ has positive value, and then
 the gauge boson may be a composite of the matter
 fermion and its anti-particle.
In the general non-abelian gauge field theory,
 the marginal value for asymptotic freedom is given by
 $N_f T(R)=11C_2(G)/4$.
Hence we find that the gauge boson can be composite
 when the fermion-flavors number $N_f$ satisfy
 the condition (\ref{COND}), in which
 the gauge field theory is not
 asymptotically free.
Therefore we find that the allowed region
 of $N_f$ in (\ref{COND})
 for the gauge boson compositeness is {\it complementary}
 to the asymptotic freedom in the gauge field theory.

When these results are applied to $SU(2)_L$ of the Glashow-Weinberg-Salam
(GWS)
 theory with twelve flavors $N_f=12$
 (three families of leptons and three families with three colors of quarks)
 and $SU(2)$ colors ($C_2(G)=N_c=2$) of their matter fermions,
 (\ref{COND}) holds, and then the weak bosons can be a composite of
 the matter fermion and its anti-particle.
If we apply the model to the quantum chromodynamics (QCD)
 with six flavors $N_f=6$ (three families of quarks)
 and $SU(3)$ colors ($C_2(G)=N_c=3$) of quarks,
 then the gluons can not be a composite of the quark
 and its anti-particle, because (\ref{COND}) does not hold.
In an application of the model to hadron physics,
 we can discuss the compositeness of light mesons.
From (\ref{COND}), the $\rho$ meson does not become composite gauge boson
 within this model,
 because the flavors number is $N_f=2$ (one family of light quark)
 though the $SU(3)$ colors is given by $C_2(G)=N_c=3$. 
In an application of the model to the system with matter fermions
 which belong to the adjoint representation of $SU(N_c)$ color gauge
 group, the gauge bosons can be a composite of the fermion
 and its anti-particle when the fermion-flavors number satisfies $N_f \geq 3$
 from (\ref{COND}).
In this case, the allowed region of $N_f$ does not depend on the colors
number $N_c$.

In the abelian gauge theory, the gauge boson can be always
 a composite of the matter fermions, because the next-to-leading order
 contribution suppressed in the theory as shown in the previous paper
 \cite{AH}.
If we apply the results to quantum electrodynamics,
 the photon may be always a composite of
 $U(1)$ charged matter fermion and its anti-particle.

In $1/N_f$ expansion, the flavors number $N_f$ need to has the large value
  in order to obtain the appropriate approximation.
In this model, when (\ref{COND}) holds,
 the coupling constant (\ref{COUPLING}) is rewritten as
\begin{eqnarray}
 g_{\rm r}^2 ={3 \over 4N_f T(R)I}
 \left[ 1 + {11C_2(G) \over 4N_f T(R)} \right],
\label{COUPLINGR}
\end{eqnarray}
 in which the second term in bracket gives the next-to-leading contribution.
For the appropriate approximation, the next-to-leading contribution in
 (\ref{COUPLINGR}) need to becomes the small.
For example, we apply (\ref{COUPLINGR}) to the system with
 matter fermions ($N_f=12$) which belong to the adjoint representation
 of $SU(2)$ color gauge group.
Then the second term in bracket of (\ref{COUPLINGR}) give the following
 value,
\begin{eqnarray}
 {11C_2(G) \over 4N_f T(R)} = {11 \over 4N_f} \approx 0.23 ,
\label{COUPLINGRn}
\end{eqnarray}
 in which the next-to-leading contribution is sufficiently small.
In this system, let us estimate the order of coupling constant of
 composite gauge field theory from (\ref{COUPLINGR}).
If we fix the scales as $\Lambda=10^{4}\ {\rm GeV}$
 and $m_{\rm r}=10^{-3}\ {\rm GeV}$, then (\ref{COUPLINGR}) give
\begin{eqnarray}
 \alpha_{\rm r} = {g_{\rm r}^2 \over 4\pi} \approx 0.03,
\label{COUPLINGRnE}
\end{eqnarray}
 in which the order of $\alpha_{\rm r}$ is equal to the order of
 $SU(2)$ weak coupling constant in the GWS theory.

\section{ CONCLUSIONS }

In order to investigate the composite gauge field,
 we have considered the compositeness condition in the general
 non-abelian gauge theory with fermionic matter fields.
The gauge field theory with
 compositeness condition
 $Z_3=0$ is equivalent to the composite gauge field theory
 where $Z_3$ is the
 wave function renormalization constant of the gauge field.
At the leading order in $1/N_f$ expansion,
 the gauge coupling constant $g_{\rm r}$ is
 proportional to
 $\sqrt{\epsilon / N_f}$ where $N_f$ is the number of
 matter-fermion flavors and $\epsilon$ is related to
 the cutoff scale $\Lambda$.
In the general non-abelian gauge field theory,
 we have calculated the renormalization constant $Z_3$
 at the next-to-leading order in $1/N_f$ expansion.
Through the compositeness condition ($Z_3=0$), we obtain
 the coupling constant in composite gauge field theory as
 $ g_{\rm r} \propto  1/ \sqrt{ 4N_f T(R) -11C_2(G) } $.
Then we have found that the gauge boson can be a composite of the matter
 fermion and its anti-particle
 when $N_f$ satisfies the relation
 $N_f >11C_2(G)/4T(R)$ because $g_{\rm r}$ has to be real,
 in which the non-abelian gauge field theory is not asymptotically free.
Therefore we find {\it the complementarity}
 that the allowed region of $N_f$ for the gauge boson
 compositeness is complementary to the asymptotic freedom
 in the non-abelian gauge field theory.
We have applied these results to the weak bosons in GWS theory
 and the gluons in QCD, and the gauge bosons in
 the system with matter fermions which
 belong to the adjoint representation of $SU(N_c)$ color gauge group.
Then the weak bosons can be a composite of the quarks and the leptons,
 but the gluons can not be a composite of the quarks.
The gauge bosons can be a composite of fermions which belong to the
 adjoint representation of $SU(N_c)$ color gauge group
 if the fermion-flavors number satisfies $N_f \geq 3$.
To the application of compositeness condition in abelian gauge theory,
 the photon in quantum electrodynamics may be always a composite of
 the $U(1)$ charged matter fermions, because the next-to-leading order
 contribution suppressed within the abelian gauge theory as shown in
 the previous paper \cite{AH}.
We have seen that when the relation $N_f >11C_2(G)/4T(R)$ holds,
 the $1/N_f$ expansion gives the appropriate approximation.
Then we can estimate the order of coupling constant of composite
 gauge field theory.

\section*{ACKNOWLEDGMENTS}
We would like to thank Professor K. Akama for helpful discussions.
We would like to thank Professor H. Terazawa for kind hospitality
 during our stay at KEK.

\epsfxsize=16.0cm\epsffile{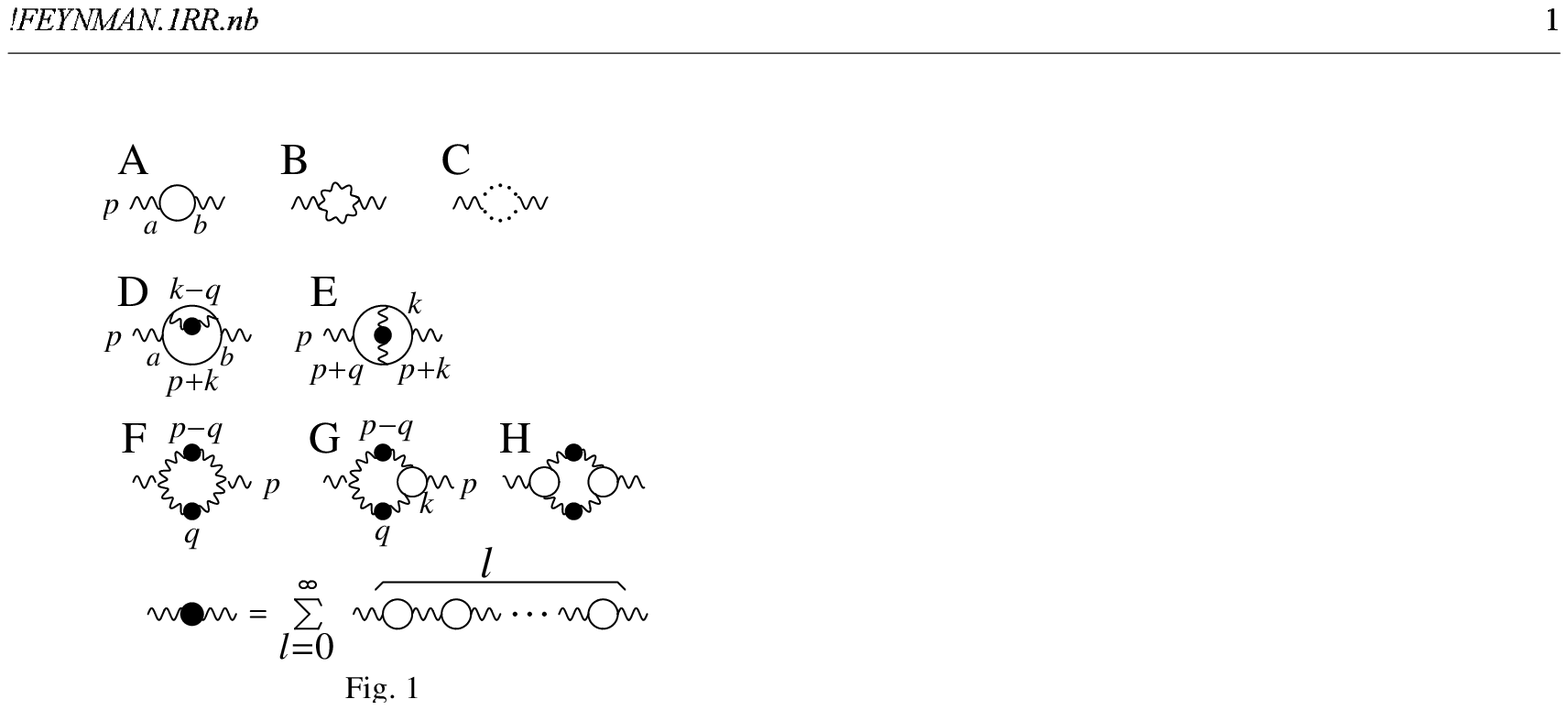}

\epsfxsize=16.0cm\epsffile{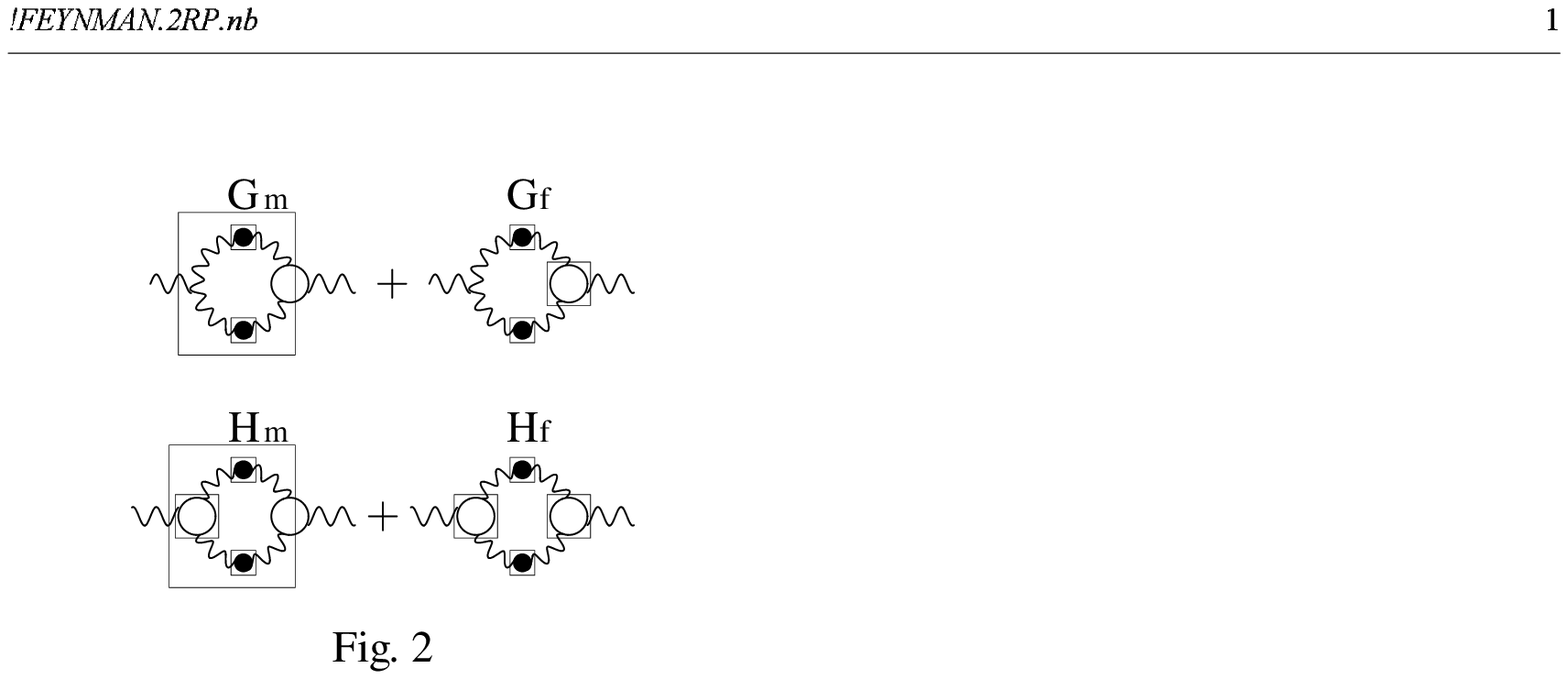}

\begin{figure}

\caption
{ The gauge boson self-energy parts at the leading order (A) and
 at the next-to-leading order (B-H) in $1/N_f$.
 The solid, wavy, and dotted lines indicate the fermion, gauge boson,
 and Fadeev-Popov ghost propagators, respectively.
 The disk stands for insertion of an arbitrary number of
 one-fermion-loops into the gauge boson propagator.
 $p$ indicates the momentum of boson in the external line.
 $q$ and $k$ are momenta in the internal line.
 $a$ and $b$ are color indices. }
\label{FigOne}
 \ 
 \ 
\caption
{ The separated overlapping divergence.
 ${\rm G_f}$, ${\rm H_f}$: the part where the divergence occurs at the
fermion loop
 subdiagram in the three boson vertex part.
 ${\rm G_m}$, ${\rm H_m}$: the part where the divergence occurs at the
boson-fermion
 (mixed) loop subdiagram in the boson-fermion vertex part. }
\label{FigTwo}

\end{figure}


\newpage



\begin{thebibliography}{99}
\bibitem{cg}     % H Bj ref AH
W.~Heisenberg,                  Rev.\ Mod.\ Phys.\ {\bf 29}, 269 (1957);
J.~D.~Bjorken,                  Ann.\ Phys.\ {\bf 24}, 174 (1963).

\bibitem{qlphys}
K.~Akama and H.~Terazawa, INS-Rep-{\bf 257} (1976);
H.~Terazawa, Y.~Chikashige and K.~Akama, {Phys.\ Rev.} {\bf D15}, 480(1977);
{Prog.\ Theor.\ Phys.} {\bf 56}, 1935 (1976) ;
T.~Saito and K.~Shigemoto, {Prog.\ Theor.\ Phys.} {\bf 57} (1977) 242;
O.~W.~Greenberg and J.~ Sucher, Phys.\ Lett.\ {\bf 99B}, 339 (1981);
H.~Terazawa, {Phys.\ Rev.} {\bf D22}, 184 (1980);
{Prog.\ Theor.\ Phys.} {\bf 64}, 1763 (1980);{\bf 79}, 734 (1988);
Mod.\ Phys.\ Lett. {\bf A13}, 2427 (1988);
L.~F.~Abbott and E.~Farhi, Phys.\ Lett.\ {\bf 101B}, 69 (1981).
M.~Yasu\`{e}, Nucl.\ Phys.\ {\bf B234}, 252 (1983);
{Prog.\ Theor.\ Phys.} {\bf 81}, 271 (1989);
{Phys.\ Rev.} {\bf D39}, 3458 (1989);{\bf D42}, 3169 (1990);

\bibitem{hadphys}
See for example, J.~J.~Sakurai, {\it Currents and Mesons},
(Univ. Chicago Press, Chicago, 1969);
M.~Bando, T.~Kugo, S.~Uehara, K.~Yamawaki and T.~Yanagida,
        Phys.\ Rev.\ Lett.\ {\bf 54} 1215 (1985).

\bibitem{cgg2}
A.~Hasenfratz and P.~Hasenfratz, Phys.\ Lett.\ {\bf B297}, 166 (1992);
M.~A.~Luty, Phys.\ Rev.\ {\bf D48}, 1295 (1993);
M.~Bando, Y.~Taniguchi and S.~Tanimura,
Prog.\ Theor.\ Phys.\ {\bf 97}, 665 (1997);
M.~Bando, J.~Sato, and K.~Yoshioka,
ICRR-Report-384-97-7, hep-ph/9703321 (1997);
M.~Bando, hep-ph/9705237 (1997).

\bibitem{cgq2}
K.~Akama and T.~Hattori,
Phys.\ Rev.\ {\bf D40}, 3688 (1989);{\bf D51}, 3895 (1995);
Int.\ J.\ Mod.\ Phys.\ {\bf A9}, 3503 (1994);
K.~Akama, T.~Hattori, and M.~Yasu\`e,
Phys.\ Rev.\ {\bf D42}, 789 (1990);{\bf D43}, 1702 (1991);
B.~S.~Balakrishna and K.~T.~Mahanthappa, Phys.\ Rev.\ {\bf D49} 2653 (1994);
Phys.\ Rev.\  {\bf D52} 2379 (1995);
A.~Galli, Phys.\ Rev.\ {\bf D51}, 3876 (1995);
          Nucl.\ Phys.\ {\bf B435}, 339 (1995);
C.~P.~Burgess, J.-P.~Derendinger, F.~Quevedo, and M.~Quiros,
Phys.\ Lett.\ {\bf B348} 428 (1995);
V.~V.~Kabachenko and  Y.~F.~Pirogov, IHEP 96-97, hep-ph/9612275 (1996);
M.~Berkooz, P.~Cho, P.~Kraus, and M.~J.~Strassler,
HUTP-97/A014, hep-th/9705003 (1997);
M.~Yasu\`e, TOKAI-HEP/TH-9701, hep-ph/9707312 (1997).


\bibitem{ccg}
B.~Jouvet,              {Nuovo Cim.} {\bf 5}, 1133 (1956);
M.~T.~Vaughn, R.~Aaron and R.~D.~Amado,
                        {Phys.\ Rev.} {\bf 124}, 1258 (1961);
A.~Salam,               {Nuovo Cim.} {\bf 25}, 224 (1962);
S.~Weinberg,            {Phys.\ Rev.} {\bf 130}, 776 (1963);
I.~Bialynicki-Birula,           Phys.\ Rev.\ {\bf 130}, 465 (1963);
D.~Luri\'e and A.~J.~Macfarlane,  Phys.\ Rev.\ {\bf 136} (1964) B816;
T.~Eguchi,  {Phys.\ Rev.} {\bf D14} (1976) 2755; {\bf D17} (1978) 611;
K.~Shizuya,  {Phys.\ Rev.} {\bf D21} (1980) 2327.

\bibitem{GNKK}
D.~J.~Gross and A.~Neveu, { Phys.\ Rev.} {\bf D10}, 3235 (1974);
T.~Kugo,                { Prog.\ Theor.\ Phys.} {\bf 55}, 2032 (1976);
T.~Kikkawa,             { Prog.\ Theor.\ Phys.} {\bf 56}, 947 (1976).

\bibitem{hid2}
M.~Tanabashi, Phys.\ Lett.\ {\bf B384}, 218 (1996);
M.~C.~Birse, Z.\ Phys.\ {\bf A355}, 231 (1996);
M.~Hashimoto, Phys.\ Rev.\ {\bf D54}, 5619 (1996);
C.~M.~Shakin and W.-D.~Sun, Phys.\ Rev.\ {\bf D55}, 2874 (1997);
L.-H.~Chan, Phys.\ Rev.\ {\bf D55}, 5362 (1997);
M.~Harada and A.~Shibata, Phys.\ Rev.\ {\bf D55}, 6716 (1997).

\bibitem{pheno}
K.~Akama and H.~Terazawa,
Phys.\ Lett.\ {\bf 112B}, 387 (1982);{\bf 321B}, 145 (1994);
Phys.\ Rev.\ D {\bf 55}, R2521 (1997);
M.~Bander, Phys.~Rev.~Lett.~\underline {77}, 601 (1996);
A.~E.~Nelson, Phys.\ Rev.\ Lett.\ {\bf 78} 4159 (1997);
S.~L.~Adler, IASSNS-HEP 97-12, hep-ph/9702378;
M.~C.~Gonzalez-Garcia and S.F. Novaes, IFT-P-024-97, hep-ph/9703346;
N.~Di~Bartolomeo and M.~Fabbrichesi, Phys.\ Lett.\ {\bf B406}, 237 (1997);
K.~Akama K.~Katsuura, and H.~Terazawa, Phys.\ Rev.\ D {\bf 56}, R2490
(1997).

\bibitem{tc}
W.A. Bardeen, C.T. Hill and M. Lindner,
                        { Phys. Rev.} {\bf D41}, 1647 (1990);
M.~Harada, Y.~Kikukawa, T.~Kugo and H.~Nakano,
                        { Prog.\ Theor.\ Phys.} {\bf 92}, 1161 (1994).

\bibitem{emb}
K.~Akama, in {Gauge Theory and Gravitation, Proceedings of
        the International Symposium, Nara, Japan, 1982}
        ed.\ K.~Kikkawa, N.~Nakanishi and H.~Nariai (Springer-Verlag)
        p.\ 267;
M.~D.~Maia, J.\ Math.\ Phys.\ {\bf 25} (1984) 2090;
K.~Akama, {Prog.\ Theor.\ Phys.} {\bf 78}, 184 (1987);
        {\bf 79}, 1299 (1988).

\bibitem{AH}
K.~Akama and T.~Hattori,
{ Phys.\ Lett.} {\bf B392}, 383 (1997);
{ Phys.\ Lett.} {\bf B445}, 106 (1998);

\bibitem{exp}
CDF Collaboration, F.~Abe {\it et al}.,
        Phys.\ Rev.\ Lett.\ {\bf 77}, 438 (1996);
        {\it ibid}.\ {\bf 77}, 5336 (1996);
        {Phys.\ Rev.\ D} {\bf 55}, R5263 (1997);
Particle Data Group, R.M.~Barnett {\it et al}.,
        {Phys.\ Rev.\ D} {\bf 54}, 1(1996);
H1 Collaboration, C.~Adloff {\it et al}.,
        Z.\ Phys.\ C.\ {\bf 74}, 191 (1997);
ZEUS Collaboration, J.~Breitweg {\it et al}.,
        Z.\ Phys.\ C.\ {\bf 74}, 207 (1997).

\bibitem{NJL}
Y.~Nambu and G.~Jona-Lasinio, {Phys.\ Rev.} {\bf 122}, 345 (1961).

\bibitem{cA}
K.~Akama, { Phys.\ Rev.\ Lett.} {\bf 76}, 184 (1996).

\bibitem{af}
D.~J.~Gross and F.~Wilczek, { Phys.\ Rev.\ Lett.} {\bf 30}, 1343 (1973);
H.~D.~Politzer, { Phys.\ Rev.\ Lett.} {\bf 30}, 1346 (1973).

\bibitem{Muta} See for example,
T.~Muta, {\it Foundations of Quantum Chromodynamics,
An Introduction to Perturbative Method in Gauge Theories},
World Scientific, (1987).

\end{thebibliography}
\end{document}